\documentclass[]{aastex631}

\shorttitle{Solid and Gaseous Methane in IRAS 23385+6053}
\shortauthors{Nakibov R. et al.}
\begin{document}

\correspondingauthor{Nakibov Ruslan}
\email{nakibov.ruslan@urfu.ru}

\title{Solid and Gaseous Methane in IRAS 23385+6053 as seen with Open JWST Data \footnote{Released on March, 1st, 2021}}

\author[0009-0004-5420-8824]{Ruslan Nakibov}
\affiliation{Research Laboratory for Astrochemistry, \\ Ural Federal University, Kuibysheva St. 48 \\
Yekaterinburg 620026, Russia}

\author[0009-0007-0109-3439]{Varvara Karteyeva}
\affiliation{Research Laboratory for Astrochemistry, \\ Ural Federal University, Kuibysheva St. 48 \\
Yekaterinburg 620026, Russia}

\author[0000-0002-1584-3620]{Igor Petrashkevich}
\affiliation{Research Laboratory for Astrochemistry, \\ Ural Federal University, Kuibysheva St. 48 \\
Yekaterinburg 620026, Russia}

\author[0000-0002-7308-9056]{Maksim Ozhiganov}
\affiliation{Research Laboratory for Astrochemistry, \\ Ural Federal University, Kuibysheva St. 48 \\
Yekaterinburg 620026, Russia}

\author[0009-0001-3921-604X]{Mikhail Medvedev}
\affiliation{Research Laboratory for Astrochemistry, \\ Ural Federal University, Kuibysheva St. 48 \\
Yekaterinburg 620026, Russia}

\author[0000-0002-0786-7307]{Anton Vasyunin}
\affiliation{Research Laboratory for Astrochemistry, \\ Ural Federal University, Kuibysheva St. 48 \\
Yekaterinburg 620026, Russia}

\begin{abstract}
We present a new description of the 7.7~$\mu$m region towards the high-mass star-forming region IRAS 23385+6053 taken from open James Webb Space Telescope Mid-Infrared Instrument Medium Resolution Spectrometer (JWST MIRI/MRS) data. This area is commonly attributed to the $\nu_4$ deformation mode of methane ice. For the first time gaseous and solid methane were analyzed simultaneously in IRAS 23385+6053. The band at 7.58--7.8 $\mu$m (1320--1280 cm$^{-1}$) is interpreted as a wide solid absorption methane feature overlapped by the sharp features of the methane emission. We report the detection of gaseous methane and estimate its emitting area radius~$R$, temperature~$T$ and column density~$N$ as $R=2940$~au, $T=103^{+13}_{-11}$~K, and $N=0.78_{+6.18}^{-0.64}\times10^{17}$~cm$^{-2}$, correspondingly. The ice content was analyzed with the laboratory spectra dataset of methane in different molecular environments obtained on the Ice Spectroscopy Experimental Aggregate (ISEAge). We were able to describe the wide feature of solid methane with the following laboratory spectra: CH$_4$~:~CO$_2$~=~1~:~5 (at $27.4^{+6.0}_{-10.8}$~K) and CH$_4$~:~H$_2$O~=~1~:~10 (at $8.4^{+16.4}_{-1.7}$~K) deposited at 6.7~K and warmed up at a rate of 0.5 K per minute. The derived column densities are $N_{\text{CH}_4}$(CO$_2$)~=~$2.97^{+0.37}_{-0.57}\times10^{17}$~cm$^{-2}$ and $N_{\text{CH}_4}$(H$_2$O)~=~$1.02^{+0.46}_{-0.27}\times10^{17}$~cm$^{-2}$. According to the best fit solid methane is mostly surrounded by CO$_2$ rather than H$_2$O. The residuals analysis reveals the unassigned region at 1283--1297~cm$^{-1}$ (7.71--7.79~$\mu$m) which is tentatively assigned to nitrous oxide (N$_2$O) in various environments.

\end{abstract}

\keywords{Laboratory astrophysics --- Ice spectroscopy --- Interstellar clouds --- Astrochemistry}

\section{Introduction} \label{sec:intro}

Methane (CH$_4$) is one of the most abundant molecules in space. It has been found both in solid and gaseous states in a large number of objects of astrochemical importance: molecular clouds \citep{lacy1991discovery, spitzer_oberg_2008,a}, protostars \citep{1998A&A...336..352B, JOYS}, comets \citep{drapatz1987comets,doi:10.1126/science.272.5266.1310}, atmospheres of planets \citep{SROMOVSKY2011292,Barman_2015}, exoplanets \citep{2008Natur.452..329S,2023Natur.623..709B}, and satellites \citep{REIDTHOMPSON1984236,mitchell2016climate}. It is one of the simplest carbonaceous molecules that is considered a precursor for more complex carbonaceous molecules \citep{2008ApJ...681.1385H,2008ApJ_CH4_chains} and plays an important role in proposed warm carbon-chain chemistry \citep{Sakai_2008}. Methane also has an astrobiological application for planetary studies \citep{long_kobayashi}, possibly serving as a biosignature \citep{2018AJ....156..114K,thompson2022case}.

For interstellar methane the first secure detection in gaseous form and possible in solid form was claimed by \cite{lacy1991discovery} and the solid methane detection was subsequently confirmed by \cite{Boogert1997}. However, methane is mostly found in solid rather than in gaseous state due to the absence of permanent electric dipole moment and radio inactivity as a consequence. Assignment of icy and gaseous features differs: in contrast to gaseous features the shape and position of solid state absorption bands in infrared spectra change depending on the molecular environment, therefore a collection of laboratory spectra with relevant components ratio is needed to correctly describe the observations. Methane ice features are consistently assigned to water-based ices (e.g.  Spitzer c2d (`Cores to Disks') Legacy program \cite{spitzer_oberg_2008}). This approach might need revisiting with the upcoming high quality James Webb Space Telescope (JWST) spectra. The future methane assignments would benefit from environment variety and accounting for different phases of methane in laboratory research \citep{Gerakines_2015,doi:10.1021/acs.jpca.9b10643}.

In this paper we focus on the 7.7 $\mu$m methane deformation mode in IRAS 23385+6053, a star-forming region recently studied by the JOYS team in \cite{JOYS}. While that paper was devoted to detection of complex organic molecules in the 6.8--8.6 $\mu$m region, we note another interesting feature. The spectrum stands out among other available JWST data and Spitzer c2d data due to its unique doublet structure. The red wing of the doublet in their Figure 6 is described by \cite{JOYS} mostly with methane and water mixture: CH$_4$~:~H$_2$O~=~1~:~10 at 15~K. Description of the blue wing (1308 cm$^{-1}$) is not provided. We also note that the `splitting' feature of the doublet is positioned at 1306 cm$^{-1}$ which is close to a $\nu_4$ gaseous methane feature \citep{1936}. Interpretation of this spectrum requires laboratory spectra of astrophysically relevant ice mixtures as well as the analysis of infrared emission of gaseous methane. In this work, we perform such analysis. Using the Ice Spectroscopy Experimental Aggregate (ISEAge) setup \citep{Ozhiganov_2024} we obtained laboratory spectra of ice mixtures containing methane that are missing in the literature. Those spectra were combined with literature data to analyze the solid-state band at 7.7~$\mu$m of the spectrum of IRAS 23385+6053. The gaseous emission features were interpreted using numerical radiative transfer simulations. This allowed us to characterize the methane contents of IRAS 23385+6053 for the first time both in the gaseous and solid phases.

\section{Observations and data reduction}

IRAS 23385+6053 (Mol160) is an object of a long-standing astrophysical interest \citep{fontani2004iras,Cesaroni2019,Beuther2023,2023A&AGieser_gasclusters,2024A&A_joys_gases}. It is located at a distance of 4.9 kpc with coordinates $\alpha$(J2000)~=~23$^h$~40$^m$~54.5$^s$, $\delta$(J2000)~=~61$^{\circ}$~10$^{\prime}$~28.1$^{\prime \prime}$ \citep{Molinari1998,Molinari2002}. Estimates of luminosity ($\sim$3000~$L_{\odot}$) obtained by \cite{Molinari2008,Cesaroni2019} and total envelope mass ($\sim$510~$M_{\odot}$) estimated by \cite{ 2018A&A...617A.100B}. IRAS 23385+6053 is a young high-mass star-forming region that has a complex structure which includes 6 dense cores \citep{Cesaroni2019}, two young stellar objects (YSO) close to each other \citep{Cesaroni2019,2024A&A_joys_gases} and three high velocity outflows \citep{Beuther2023}. The material toward the IRAS 23385+6053 consists of relatively cold gas and dust ($\sim$50~K), along with warm gas ($\sim$400~K) created by outflows as shown in \cite{2023A&AGieser_gasclusters}.

\begin{figure}[!ht]
    \centering
    \includegraphics[width=\textwidth]{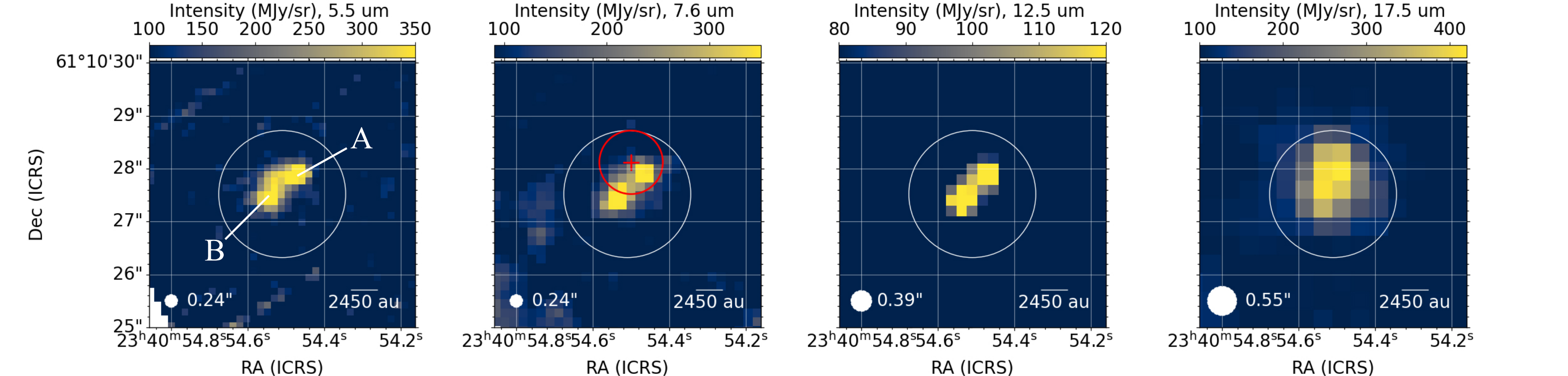}
    \caption{Intensity maps of IRAS 23385+6053 at different wavelengths from the first, second, third and fourth channels of JWST MIRI-MRS spectrograph. The white circle shows 2.4$^{\prime\prime}$ aperture from which the spectrum is extracted. Channels have different pixel size and map size depending on the observed wavelength. The young stellar objects are designated as A and B as in \cite{2024A&A_joys_gases}. The white circle in the lower left corner shows the FWHM. The red circle shows the surface emission area of gaseous methane. The red cross shows the methane gas emission peak.}
    \label{Observ}
\end{figure}

\begin{figure}[!ht]
    \centering
    \includegraphics[width=\textwidth]{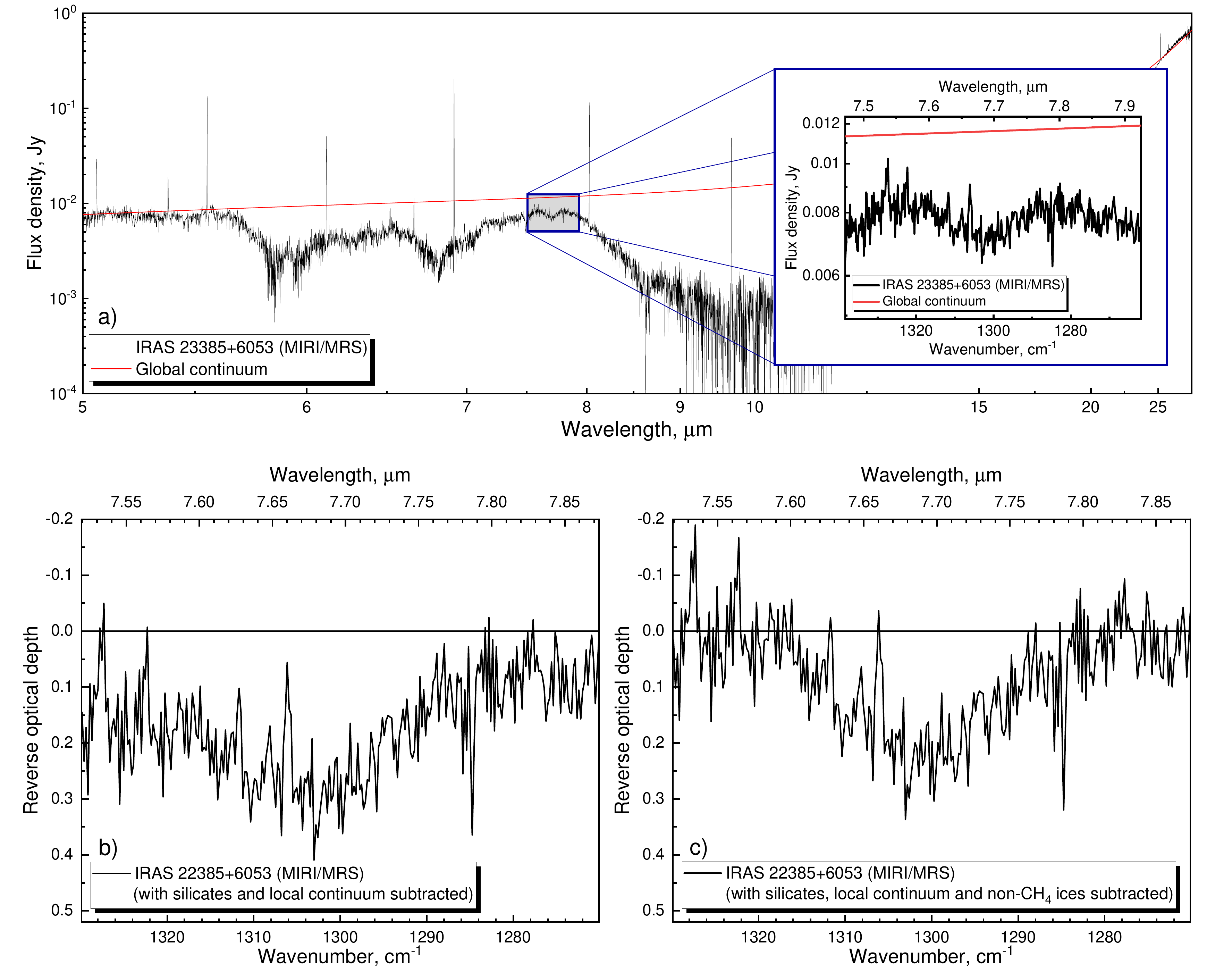}
    \caption{Panel \textit{a} --- The IRAS 23385+6053 spectrum after background subtraction. Global continuum is adopted from \cite{JOYS}. The inset shows a zoom in on the 7.7 $\mu$m region. Panel \textit{b} --- A processed spectrum after global, local continuum and a silicate profile subtraction. Panel \textit{c} --- Same as \textit{b}, but with non-CH$_4$ ices (OCN$^-$, CH$_3$CH$_2$OH, SO$_2$) subtracted following \cite{JOYS}}
    \label{continuum}
\end{figure}

JWST observations towards IRAS 23385+6053 were first presented in \citep{Beuther2023}, which showed the presence of two mid-infrared continuum sources. Observations were taken from the MAST database (MAST:\dataset[10.17909/qv13-yt46]{http://dx.doi.org/10.17909/qv13-yt46})  with the level 3 pipeline calibration and were obtained at JWST with Mid-IR Instrument (MIRI) Medium Resolution Spectrograph (MRS). Fig. \ref{Observ} shows that both sources are resolved in the short-wave part of the spectrum (channels 1 and~2, $\lambda<12$~$\mu$m) and becoming unresolved in the long-wave part of the spectrum (channels 3 and 4, $\lambda>12$~$\mu$m). We extracted the spectrum from a cylindrical aperture with a diameter of 2.4$^{\prime\prime}$ centered in the direction between the sources (J2000 RA: 23$^h$ 40$^m$ 54.51$^s$, DEC: 61$^{\circ}$ 10$^{\prime}$ 27.52$^{\prime \prime}$, see Fig. \ref{Observ}) to cover the emission from sources in all channels. In \cite{JOYS} it is noted that additional background observations are not used to consider background because other sources are included in the background observations. Therefore, we estimated the background by extracting the spectrum towards the region of lowest background flux outside the main infrared sources \citep[J2000 RA: 23$^h$ 40$^m$ 54.15$^s$, DEC: 61$^{\circ}$ 10$^{\prime}$ 26.96$^{\prime \prime}$,][]{JOYS}. The spectrum after background subtraction is shown in Fig. \ref{continuum}.

The data obtained from MAST was processed following the steps described in \cite{JOYS}: we adopted the global continuum given by fifth-order polynomial, the mixtures combination of olivine and pyroxene with carbon (87\%~:~13\%) produced by \texttt{optool} code \citep{2021ascl.soft04010D} were used for silicate subtraction. We further subtract the local continuum utilizing the third-order polynomial with guiding points placed at 6.80, 7.15, 7.80, and 8.42~$\mu$m.
No smoothing or rebinning was performed in order to preserve the sharp gaseous features. 

\section{EXPERIMENTAL METHODS} 

In this work the Ural Federal University  ISEAge setup is utilized. The detailed information about the setup can be found in \cite{Ozhiganov_2024}. Briefly, the setup allows for an ice growth of the controlled composition under ultrahigh vacuum conditions and cryogenic temperatures. The main chamber is continuously evacuated by the means of turbomolecular pump in combination with a forevacuum pump allowing for a base pressure of $2\cdot10^{-10}$~mbar prior to deposition (measured by Pfeiffer Vacuum IKR 270). The gaseous material is introduced from all-metal dosing line through leak valves. The components ratio is monitored using a Stanford Research Systems RGA200 quadrupole mass spectrometer. The ion currents for the mass-spectroscopy are chosen to reflect the desired concentration as well as to account for nonlinear dependency of ice growth rate on ion currents according to protocol similar to used in \cite{Slavicinska_et_al._(2023)}. The gaseous material is condensed on Ge window, surrounded by custom made gold-coated radiation shield. The window can be used in a temperature range from 6.5~K to 305~K controlled by Lakeshore 335 controller through the application of the resistive (50~$\Omega$) heating in combination with LakeShore DT-670B-SD standard curve Silicon diode used for monitoring. The Ge window is cooled by closed-cycle helium cryostat. The symmetrical geometry of the setup leads to equal deposition of introduced material on both sides of Ge window utilizing the `background deposition' method. Transmittance IR spectra then are acquired using the Thermo Scientific Nicolet iS50 FTIR spectrometer in the range between 4000 and 630~cm$^{-1}$ (2.5 and 15.9~$\mu$m) with 1~cm$^{-1}$ resolution. FTIR spectrometer and the whole path of the IR beam outside of the UHV main chamber is flushed with pure N$_2$ to minimize atmospheric gas contamination of the spectra.

The compounds used are as follows: CO (99.9999~\%, Ugra-PGS), CO$_2$ (99.9999~\%, Ugra-PGS),  deionized H$_2$O, CH$_4$ (99.999~\%, Ugra-PGS). Gaseous CO, CO$_2$ and CH$_4$ are introduced directly into the dosing lines from the commercially acquired gas bottles. H$_2$O was purified by a freeze-pump-thaw cycle three times before each experiment.

\section{Laboratory data}

Given the limited amount of methane spectra available in the literature, and taking into account the temperature gradient in protostars of IRAS 23385+6053 we conducted our own series of experiments. The ices were warmed to mimic the temperature changes of the interstellar dust grains during the protostar evolution. The following mixtures were obtained and warmed up: binary ices of CH$_4$ with H$_2$O, CO$_2$, NH$_3$, CH$_3$OH. These molecules are the main constituents of the polar layer of ice on a surface of the dust grain where methane is assumed to form \cite{1997A&A...317..929B}. The ratios of ice components were chosen based on astrochemical predictions for low-mass and massive YSOs (see Table 2 in \cite{boogert2015observations_Ch4}). 

\begin{figure}[!ht]
    \centering
    \includegraphics[width=0.7\textwidth]{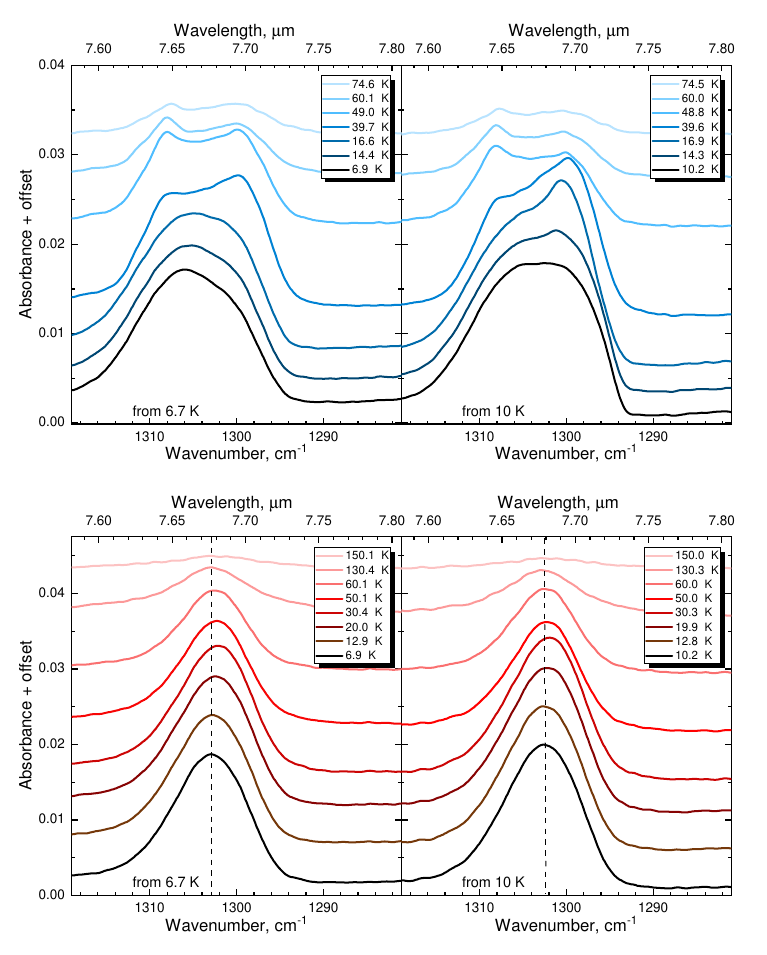}
    \caption{Top panel --- the warm up of the mixture CH$_4$ : CO$_2$ = 1 : 5  from 6.7 K and 10 K with 0.5 K per minute; bottom panel --- the warm up of the mixture CH$_4$ : H$_2$O = 1 : 10  from 6.7 K and 10 K with 0.5 K per minute. The CO$_2$ mixture is seen to be strongly influenced by the deposition temperature. The dotted lines mark the peak positions for H$_2$O mixtures observed before the warm up and serve as visual guides.}
    \label{lab_specs}
\end{figure}

We performed two sets of experiments with the deposition  temperature of 6.7~and 10~K, which correspond to temperatures of pure crystalline II* and II phases of solid methane \citep{doi:10.1021/acs.jpca.9b10643}, and thermally programmed desorption (TPD) afterwards (at a rewarmth rate of 0.5~K per minute) continuously recording both IR and mass-spectra. Results of these experiments will be thoroughly described in a separate follow-up paper. Here we focus on the mixtures utilized in this paper: CH$_4$~:~H$_2$O~=~1~:~10, CH$_4$~:~CO$_2$~=~1~:~5 (Fig. \ref{lab_specs}).

The spectra of all mixtures were obtained following the procedure described below. A leak valve containing gaseous methane was opened at first and shortly after reaching desired ion current a second leak valve was opened thus starting the mixture deposition. The deposition rate of methane onto a Ge substrate both at 6.7 and 10~K remained constant at $V = 5.9 \times 10^{12}$ cm$^{-2} $s$^{-1}$ and the deposition continued for 120 minutes. Thus, we get a column density of $4.25\times 10^{16}$~cm$^{-2}$ for CH$_4$. Extra precautions were made to ensure the same deposition rate for methane depositions in phases II and II* as those have different band strengths (A(10~K)~=~$8.4\times10^{-18}$~cm~molecule$^{-1}$ from \cite{bouilloud2015bibliographic} for peak position 1301~cm$^{-1}$,  A(8~K)~=~$1.04\times10^{-17}$~cm~molecule$^{-1}$ from \cite{Gerakines_2015} for peak position 1297~cm$^{-1}$), we used different calibrated ion currents for different temperatures. During the deposition and TPD the infrared and mass-spectra were recorded. The IR spectra were recorded every 45 seconds (an average of 32 scans) with a single spectrum averaging 128 scans at the end of each deposition.

On obtaining IR spectra of CH$_4$~:~CO$_2$ mixture with the ratio of 1~:~5 at 10~K with a rewarmth rate of 0.5~K per minute a doublet with peak positions matching the observational data is observed after $\sim$40~K. It is also the only mixture that covers both wings in width. We attribute this to differences in ice structure upon deposition. We note that \cite{1997A&A...317..929B} mentions the doublet structure of methane ice in the CO$_2$ matrix during the warm up of CH$_4$~:~CO$_2$ mixture with a rate of 2~K per minute, however, no reference spectra are provided in that paper.

The observed width and shape ultimately led us to choose a CH$_4$~:~CO$_2$~=~1~:~5 as one of the main constituents for fitting. Another component is the CH$_4$~:~H$_2$O~=~1~:~10 mixture chosen due to ubiquity of water in icy mantles. 

\section{spectral fitting procedure}
 
We assume the resolved doublet band to be a combination of wide ice band with a sharp 1306 cm$^{-1}$ emission feature. While being considered a volatile specie gaseous methane is rarely detected in the gas phase due to the absence of a permanent electric dipole moment. IRAS 23385+6053 displays a temperature gradient on the chosen line of sight \citep{2023A&AGieser_gasclusters,2024A&A_joys_gases}. Therefore, on lines of sight towards protostars in the object, there are regions with various gas and dust temperatures. Thus, it is reasonable to assume that observed methane spectra can be interpreted using warmed-up laboratory ice spectra. Next, it is natural to explore the observational spectrum for the signs of gaseous methane. Indeed, some of the gaseous species are already analyzed in \cite{2024A&A_joys_gases}, however, methane is not covered.

In order to fit gaseous and solid phases simultaneously we utilized a two-step procedure given the differences in techniques used for different phases. Firstly we estimated the temperature and column density of gaseous methane. The careful statistical treatment is required, since gaseous features on the spectrum considered in this study have low signal-to-noise (S/N) ratio. Therefore, the more common approach of estimating ice continuum by a spline with guiding points placed at emission-free regions was discarded. On the flux density spectrum the prominent gaseous features and heavy outliers of unknown origin were masked and the spectrum itself was smoothed (locally estimated scatterplot smoothing (LOESS, \cite{loess}), 20 points window). The result, representing ice continuum, was subtracted and the residuals were tested for normality, assuming normal distribution. 

Following the research on gaseous species of IRAS 23385+6053 \citep{2024A&A_joys_gases} a simple LTE slab model was used to emulate emission of gaseous methane, the flux density $F$ is given as 
\begin{equation}\label{lte}
    F(\lambda)=\Omega B_\nu(T)(1-e^{-\tau})=\frac{S}{d^2} B_\nu(T)(1-e^{-\tau})\approx\frac{S}{d^2} B_\nu(T)k(\nu,T)N,
\end{equation}
where $\Omega=S/d^2$ is the solid angle subtended by the aperture, $S$ is the area of emitting surface, $d=4.9$ kpc is the distance to the object, $B_\nu$ is the Plank's law, $\tau$ is the optical depth, $N$ is the column density, $k(\nu,T)$ is the absorption coefficient and $\tau = kN$. For this study linearization introduces an additional 5\% error. One cannot estimate $S$ or $N$ independently from this formula, instead estimating the $S\cdot N$ value and temperature $T$, which controls the relative intensity of bands. 

The area and column density were recovered by estimating the area $S$ directly from the flux intensity maps. The upper area limit is given as 2.4$^{\prime\prime}$ from which the original spectrum was extracted. Since gaseous methane is distributed across this aperture we investigated each pixel to analyze gaseous methane distribution using the S/N ratio. It was found that gaseous methane features have low S/N ratio in single pixels ($\leqslant 2.3$) and therefore we decided to integrate the signal using aperture extraction centered at the pixel with strongest gaseous methane emission. Setting the significant S/N ratio to 3 we estimate the lower emitting area limit as the 1.2 arcsec aperture (5880 au diameter). 

To estimate the temperature and column density the $\chi^2$ was minimized, given as
\begin{equation}
    \chi^2=\sum^{N_{obs}}_{i=1}\frac{(F_{obs}^i-F_{model}^i)^2}{\sigma^2},
\end{equation}
where $N_{obs}$ is the number of analyzed data points, $F_{obs}$ and $F_{model}$ are observational and modeled flux, correspondingly, and $\sigma$ is the standard deviation of the observational spectrum. The confidence intervals were estimated from $\Delta \chi^2_{red}$-maps following \cite{avni}, with $\chi^2_{red}=\chi^2/df$ and $df=1$ degrees of freedom. The methane emission spectra were modeled for temperatures between 10 and~150 K with a 1 K step using the TheoReTS 80 K methane line lists and partition function by \cite{methane_lines} \footnote{URL: \url{https://theorets.tsu.ru/molecules.ch4.low-room-t}}. The spectra were Doppler shifted by -50.2 km~s$^{-1}$ \citep{2018A&A...617A.100B} and convolved to an average ($R\approx3500$) JWST MIRI-MRS instrument resolution \citep{Labiano}.

For the ice fit we calculated optical depth using the global continuum and subtracted the local continuum to isolate the ice features. For $\chi^2$ analysis the noise variance $\sigma^2$ was extracted following the procedure for the gaseous fit. Given that there is no standard distribution for the logarithmic transformation (flux to optical depth) we assume normal distribution. We used the linear combinations of laboratory spectra (CH$_4$~:~H$_2$O, CH$_4$~:~CO$_2$) at different temperatures $T$ for fitting with the scale coefficients of the linear combination being estimated from the fit. Therefore, for each pair of spectra the column densities $N_{\text{CH}_4 \text{ in H}_2\text{O}}$ and $N_{\text{CH}_4 \text{ in CO}_2}$ were estimated. To obtain the scale coefficients $\chi^2$ was minimized, given as
\begin{equation}
    \chi^2=\sum^{N_{obs}}_{i=1}\frac{(\tau_{obs}^i-\tau_{model}^i)^2}{\sigma^2},
\end{equation}
where $\tau_{obs}$ and $\tau_{model}$ are observational and modeled optical depths, correspondingly, and $\sigma$ is the standard deviation of the observational spectrum. In addition to minimizing $\chi^2$, the residuals were tested for normality (see Appendix~\ref{chi} for details).

Following the fitting performed in \cite{JOYS} we also added to the fit the cyanate ion (12~K OCN$^{-}$, formed in HNCO~:~NH$_3$~=~1~:~1), CH$_3$CH$_2$OH~:~H$_2$O, and SO$_2$~:~CH$_3$OH spectra obtained in \cite{2001A&A...379..588N}, \cite{spirt} and \cite{1997A&A...317..929B}, correspondingly, taken from the Leiden Ice Database for Astrochemistry (LIDA) since those describe well the wide component superimposed on the blue wing of the target doublet. We discard the need of accounting for bands distortion due to grain shape pointed by \cite{dartois2024spectroscopic} since the 7.69~$\mu$m (1300~cm$^{-1}$) methane band is more sensitive to changes in its environment rather than to a variability of the grain size: the matrix and temperature effects should prevail provided the CH$_4$ concentration does not exceed 30\% of the whole matrix \citep{1997A&A...317..929B}. 

\section{Results and discussion}
\begin{figure}[!ht]
    \centering
    \includegraphics[width=\textwidth]{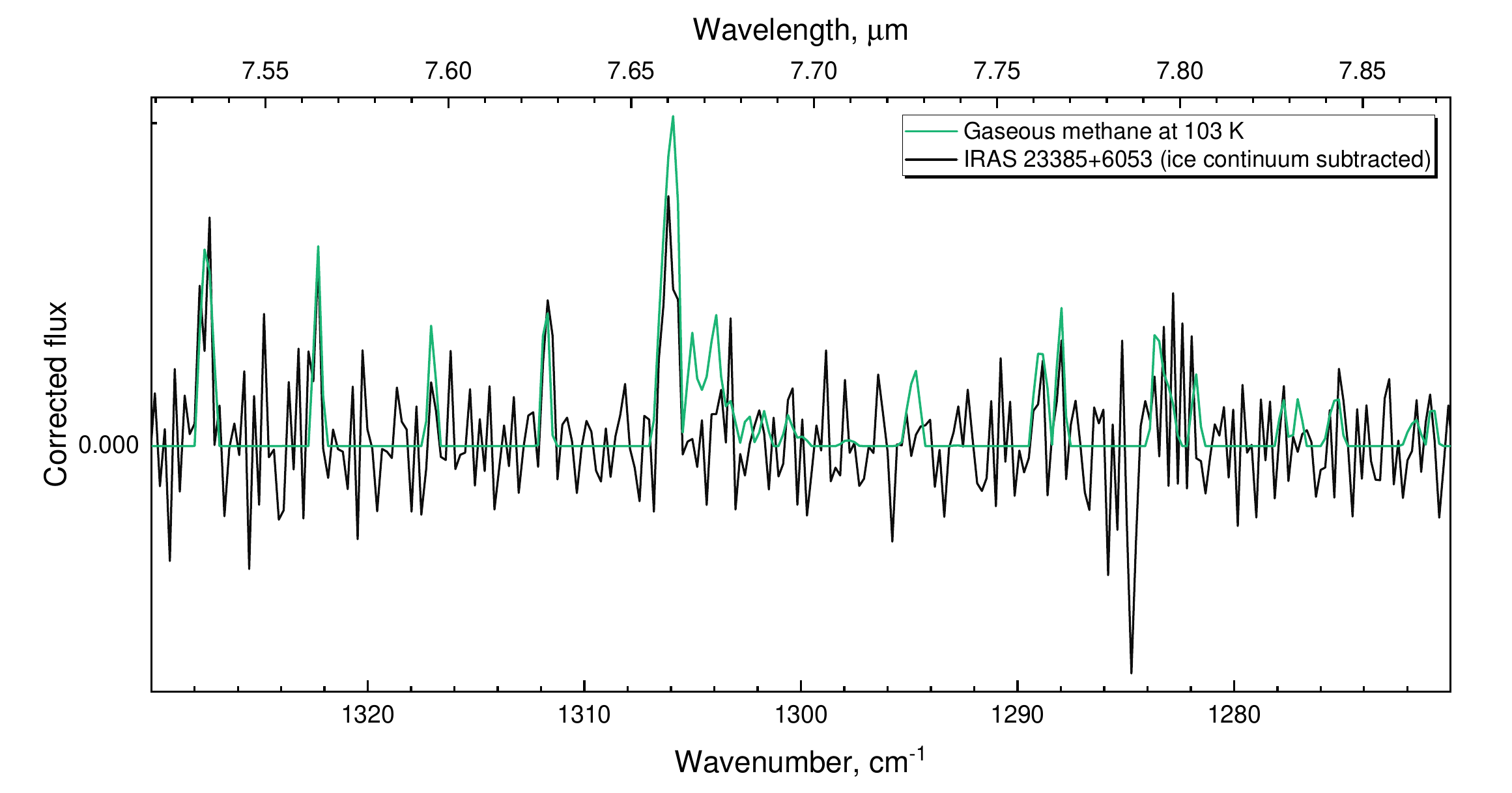}
    \caption{Best simulated gaseous methane spectrum plotted against observational data with subtracted ice continuum. A~good agreement is shown between the prominent gaseous features and outlying sharp features in the observational spectrum.}
    \label{lines}
\end{figure}

Following the procedures described in sections above, we derive the emitting area, temperature and column density of gaseous methane to be $S = 2.7\times10^7$~au$^2$ ($R=2940$~au), $T = 103^{+13}_{-11}$ K (1$\sigma$ level) and $N = 0.78_{+6.18}^{-0.64}\times10^{17}$~cm$^{-2}$, correspondingly. The best fit of the gaseous methane spectrum is shown on see Fig.~\ref{lines}. The protostar has a heated core and displays a temperature gradient up to 100~K in cold regions as shown in \cite{2023A&AGieser_gasclusters}. Other paper by \cite{2024A&A_joys_gases} concerning gas-phase molecules suggests dust temperatures above the CO$_2$ sublimation temperature closer to protostars. The intensity maps analysis shows that the signal from the most prominent gaseous feature (1306 cm$^{-1}$, S/N ratio of 2.3) does not match the position of the source emission peak, instead being offset from it (see red aperture in Fig. \ref{Observ}). The high best-fit temperature of gaseous methane implies that the emitting methane is located only in the most massive source (source~A) with a supermassive accretion disk. Based on the fitted temperature, we assume that the emitting methane is located at the edge of the accretion disk or in the envelope.

\begin{figure}[t]
    \centering
    \includegraphics[width=0.95\textwidth]{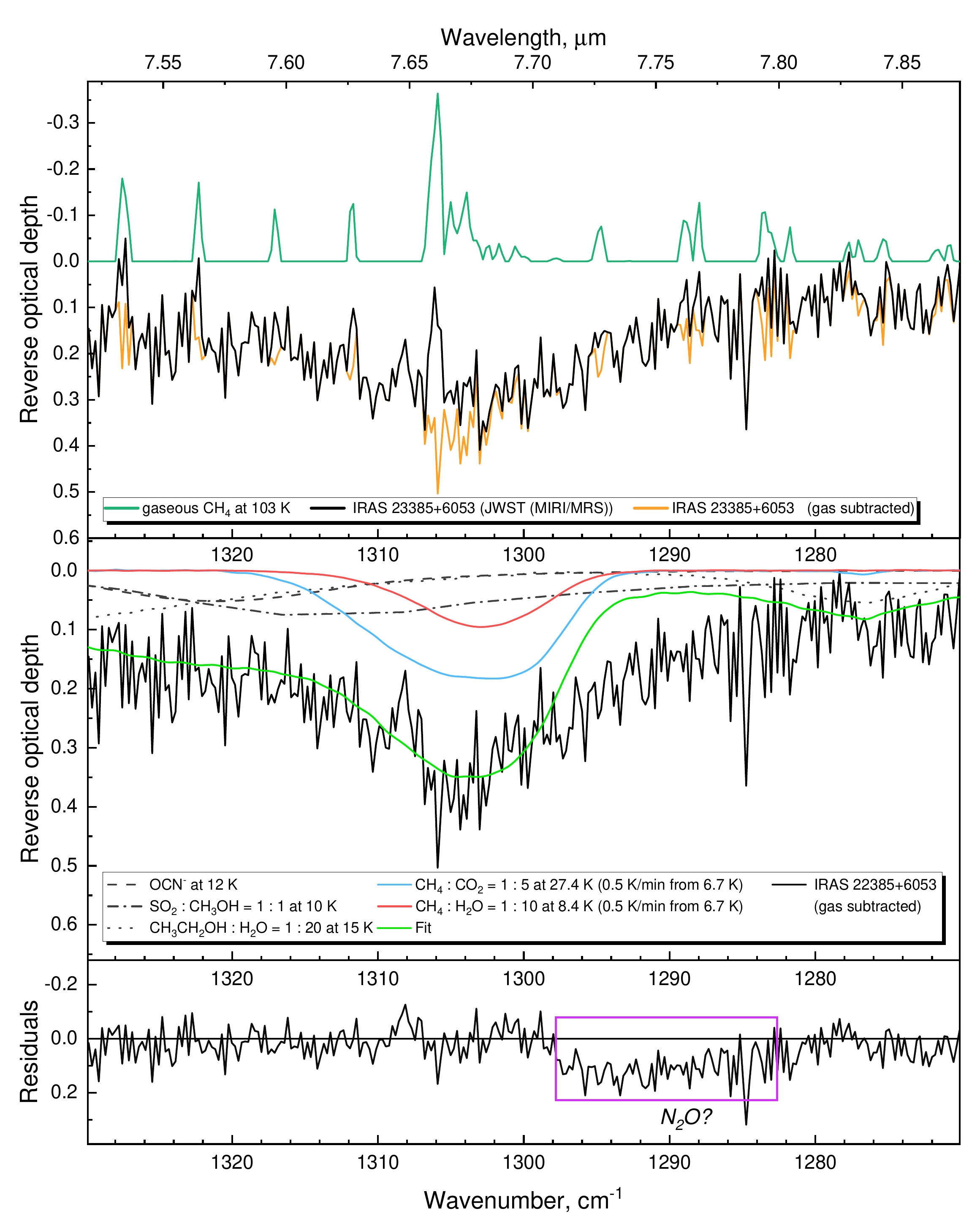}
    \caption{Observational data fitted with ISEAge laboratory mixtures (solid lines), LIDA data (dashed/dotted lines), and simulated methane emission spectrum derived from TheoReTS data.}
    \label{fit}
\end{figure}

The subtraction of fitted gaseous features reveals an unsplitted broad ice feature spanning from $\sim$1320~cm$^{-1}$ to $\sim$1280~cm$^{-1}$ (7.58--7.8 $\mu$m), see Fig. \ref{fit}, top panel. The analysis shows that the whole span cannot be described by neither the solid methane in various environments nor by a complex set of COMs, as can be seen in Figure 6 in \cite{JOYS}. Following the described procedure the CH$_4$~:~CO$_2$ and CH$_4$~:~H$_2$O spectra of different temperatures were paired and their linear combination fitted to the IRAS 23385+6053 spectrum.

In the Fig. \ref{fit} one can see a fit of the observational data with laboratory IR spectra and the simulated gaseous spectrum. The `doublet' is fully described by mixtures with water and carbon dioxide overlapped by the emission spectrum of gaseous methane. Most methane budget is best described by the CH$_4$~:~CO$_2$ mixture (1~:~5 at 27.4~K heated from 6.7~K, 0.5~K per minute) with an addition of CH$_4$~:~H$_2$O mixture (1~:~10 at 8.4~K heated from 6.7~K, 0.5~K per minute). This falls in contrast with the methane-water mixtures used predominantly for the description of this feature. However, it is to be duly noted that the amount of high quality spectra displaying the finer spectral details is limited. Further JWST data will show whether this result reflects unique properties of the protostar or is a common pattern. This result also points at the importance of accounting for ices grown at the temperatures corresponding to methane II* phase. It is also worth mentioning that \cite{spitzer_oberg_2008} reports a `curious' correlation between column densities of methane and CO$_2$, which is stronger than correlation between methane and H$_2$O, although the former weakens if only the objects with known CO abundances are considered. The strong spatial correlation between methane and CO$_2$ is reported in \cite{lei2024spatialdistributionrmch4} for IRAS 16253-2429, but not for IRAS 23385+6053.

The estimated values for temperatures and abundances are presented in Table~\ref{1}. The temperature of CH$_4$~:~CO$_2$ ice is well-constrained at $T=27.4^{+6.0}_{-10.8}$ K (1$\sigma$ level) and the temperature for CH$_4$~:~H$_2$O is estimated as $T=8.4^{+16.4}_{-1.7}$ K. The lower limit for the CH$_4$~:~H$_2$O mixture is the deposition temperature of 6.7 K. 

We performed calculations to obtain the relative absorption band strength of methane in mixtures with water and carbon dioxide in order to get more accurate column density estimations. We estimate the best-fit column density of methane in the H$_2$O environment for IRAS 23385+6053 as $N_{\text{CH}_4}$(H$_2$O)~=~$1.02^{+0.46}_{-0.27}\times10^{17}$~cm$^{-2}$ and the column density for methane surrounded by CO$_2$ as $N_{\text{CH}_4}$(CO$_2$)~=~$2.97^{+0.37}_{-0.57}\times10^{17}$~cm$^{-2}$. In view of the above, we estimate the total methane column density on the line of sight to be $N_{ice (CH_4)}=3.99\times10^{17}$~cm$^{-2}$  ($N_{ice (CH_4)}/N_{ice (H_2O)} = 2.44\%$), which matches the value from \cite{JOYS} ($N_{ice (CH_4)}/N_{ice (H_2O)} = 2.72\%$) within the margin of error. The calculated absorption band strengths are shown in the Table~\ref{1}. The column density of water on the line of sight for IRAS 23385+6053 was taken from \cite{JOYS}. 

As seen from the residuals graph only the region around 1290~cm$^{-1}$ (7.75 $\mu$m) of the whole span is left unassigned. We tentatively attribute the $\sim$1297--1283 cm$^{-1}$ (7.71--7.79 $\mu$m) region to the nitrous oxide (N$_2$O) after the additional experiments with N$_2$O-bearing ices on the ISEAge setup. The 1283 cm$^{-1}$ feature of pure nitrous oxide shifts to the blue in mixtures, reaching the largest shift with carbon dioxide (N$_2$O~:~CO$_2$~=~1~:~20). There, the N$_2$O feature shifts by 14.4 cm$^{-1}$ to as far as 1297.4 cm$^{-1}$. The spectra of pure N$_2$O and in mixtures with H$_2$O, CO$_2$, and CO are presented in Appendix \ref{aaa}. The secure detection, however, is hindered by the lack of data on the main N$_2$O feature at 2221.5 cm$^{-1}$ (4.5 $\mu$m). The same underfitting can be seen in the object NGC 1333 IRAS 2A (Figure~6 in \cite{JOYS}) in the region presumably belonging to nitrous oxide. IRAS 23385+6053 is also considered in that paper but the assumed N$_2$O region is fitted well. The N$_2$O is also tentatively detected in pure form in 4.5 $\mu$m region \citep{nazari2024hunting} and, irradiated, in 7.7 $\mu$m region \citep{rocha2024iceinventoryprotostarced}.

\begin{deluxetable*}{cccccc}
\tablenum{1}\label{1}
\tablecaption{The list of derived column densities and absorption band strengths for the deformation mode of methane ($\nu_4$)\label{tab:table1} at 7.7 $\mu$m} 
\tablewidth{0pt}
\tablehead{
\colhead{solid CH$_4$} & \colhead{Temperature} & \colhead{Band Strength} & \colhead{References}& \colhead{IRAS 23385+6053} & \colhead{IRAS 23385+6053}\\
\colhead{}       & \colhead{$T$, K}       & \colhead{$A$, cm}       & \colhead{}    & \colhead{$T$, K}          & \colhead{$N_{ice (CH_4)}\times10^{17}$, ~cm$^{-2}$}
}
\startdata
pure        & 8   & 1.04\texttimes10$^{-17}$ & \cite{Gerakines_2015} & $-$             & $-$                    \\
with CO$_2$ & 6.7 & 1.0\texttimes10$^{-17}$  & This work             & $27.4^{+6.0}_{-10.8}$      & $2.97^{+0.37}_{-0.57}$ \\1:5     \\
with H$_2$O & 6.7 & 8.9\texttimes10$^{-18}$  & This work             & $8.4^{+16.4}_{-1.7^\star}$ & $1.02^{+0.46}_{-0.27}$ \\1:10    \\
\hline
gaseous CH$_4$ & Emitting area radius  &  & Reference & IRAS 23385+6053    & IRAS 23385+6053                              \\
               & $R$, au               &  &           & $T$, K             &  $N_{gas (CH_4)}\times10^{17}$, ~cm$^{-2}$   \\
\hline
               & 2940                  &  & This work & $103^{+13}_{-11}$  & $0.78_{+6.18}^{-0.64}$                       \\
\hline
\enddata
\tablecomments{$^\star$ The lower limit for CH$_4$~:~H$_2$O mixture is the deposition temperature: 6.7 K.}
\end{deluxetable*}

\section{Conclusions} 

In this work the 7.7 $\mu$m band of IRAS 23385+6053 was analyzed based on the open JWST data. The object is a complex protostar that has a temperature gradient on the line of sight. The 7.7 $\mu$m band contains overlapped features of both icy and gaseous methane. We fit an emission spectra of gaseous methane before the solid phase fit, confirming the detection by multiple matches between the simulated and observed spectra. The fitting for the ice composition was performed with the mixtures obtained with ISEAge setup for CH$_4$-bearing ices deposited at 6.7~K and 10~K which correspond to amorphous and crystalline phases of methane. In contrary to water-methane ices being predominant in literature we show that an investigation of other binary mixtures, their deposition temperature and thermal evolution is necessary for interpretation of observational data on methane in IRAS 23385+6053. The results of this work can be summarized as follows:
\begin{enumerate}
    \item The gaseous methane emission was detected in IRAS 23385+6053, with its temperature and column density estimated ($T = 103$ K, $N = 0.78\times 10^{17}$~cm$^{-2}$). The emission is located only near the most massive source (source A, see Fig. \ref{Observ}) likely on the edge of the accretion disk or in the envelope.
    \item A new assignment is found for the 7.7 $\mu$m CH$_4$ deformation mode using our new laboratory ice mixtures CH$_4$~:~H$_2$O~=~1~:~10 (8.4 K) and CH$_4$~:~CO$_2$~=~1~:~5 (27.4 K). The column density of methane for this object is $3.99\times10^{17}$~cm$^{-2}$.  
    \item We claim tentative detection of nitrous oxide (N$_2$O). The 7.71--7.79 $\mu$m ($\sim$1297--1283 cm$^{-1}$) region is underfitted in both our research, which was specialized on various methane mixtures, and JOYS team paper focused on complex organic molecules (COMs). Our laboratory findings show that various nitrous oxide mixtures fill this region. However, the data on a more prominent N$_2$O feature is lacking for now rendering us unable to claim a secure detection.
\end{enumerate}

For the first time, the methane content of IRAS 23385+6053 was quantified simultaneously both in the gas phase and icy mantles of interstellar grains using the publicly available IR spectrum recently obtained with JWST. With the new high quality JWST spectra to come in this Letter we show a route for more secure methane detection on spectra via new spectral data and advanced statistical analysis. By this Letter we would like to emphasize the importance of the deposition temperature on CH$_4$ laboratory spectra. The importance of including amorphous methane was stated before by \cite{Gerakines_2015}. The mixtures deposited at 6.7 K being best fits shows that methane may form at the temperatures below 9 K and then get warmed up from the amorphous state. The collected IR spectra will be released in a further publication concerning methane bearing ices (Karteyeva, in prep.).  

\begin{acknowledgments}
    This research work is funded by the Russian Science Foundation via 23-12-00315 agreement. We express our gratitude to Dr. Gleb Fedoseev for sharing his expert knowledge in laboratory astrochemistry. We thank the anonymous reviewer for their insightful comments that helped us to improve the manuscript.
\end{acknowledgments}

\appendix
\section{Appendix}\label{chi}
The $\chi^2$ is a somewhat `global' statistic that is not sensitive to residuals distribution, and a single variance value output may lead to surprising results (see e.g. Anscombe's quartet, \cite{Anscombe}). Also, some of the observed features might not be present in the laboratory spectra datasets. In that case the $\chi^2$ statistic is prone to producing outliers of both signs in the residuals, diverging their distribution from normal. We find the Anderson-Darling test \citep{AD_1952,stephens_1974} to be applicable since it is a powerful normality test \citep{ad_2nd} that places attention on tails of the distribution. The $\chi^2$ for fitting was minimised via \texttt{scipy.optimize.minimize} and the Anderson-Darling test was performed via \texttt{scipy.stats.goodness\_of\_fit} afterwards \citep{2020SciPy-NMeth}. For this fit the distribution was centered at zero while the scale was fitted. 

The application of the normality test serves as a filtering step for the fits with deviation from normality in the residuals. To derive the best fit values the fits were filtered with sensitive to deviations from normality $\alpha=12\%$ ($p>0.12$) confidence level, and for estimating the confidence regions the fits were filtered with a more conservative $\alpha=10\%$ ($p>0.10$). The confidence intervals were estimated from $\Delta \chi^2_{red}$-maps following \cite{avni}, with $\chi^2_{red}=\chi^2/df$ and $df=2$ degrees of freedom. This way we ensure robustness in the best-fit point while still capturing broader variability in the confidence region.

\section{Appendix}\label{aaa}
\begin{figure}[h]
    \centering
    \includegraphics[width=0.75\textwidth]{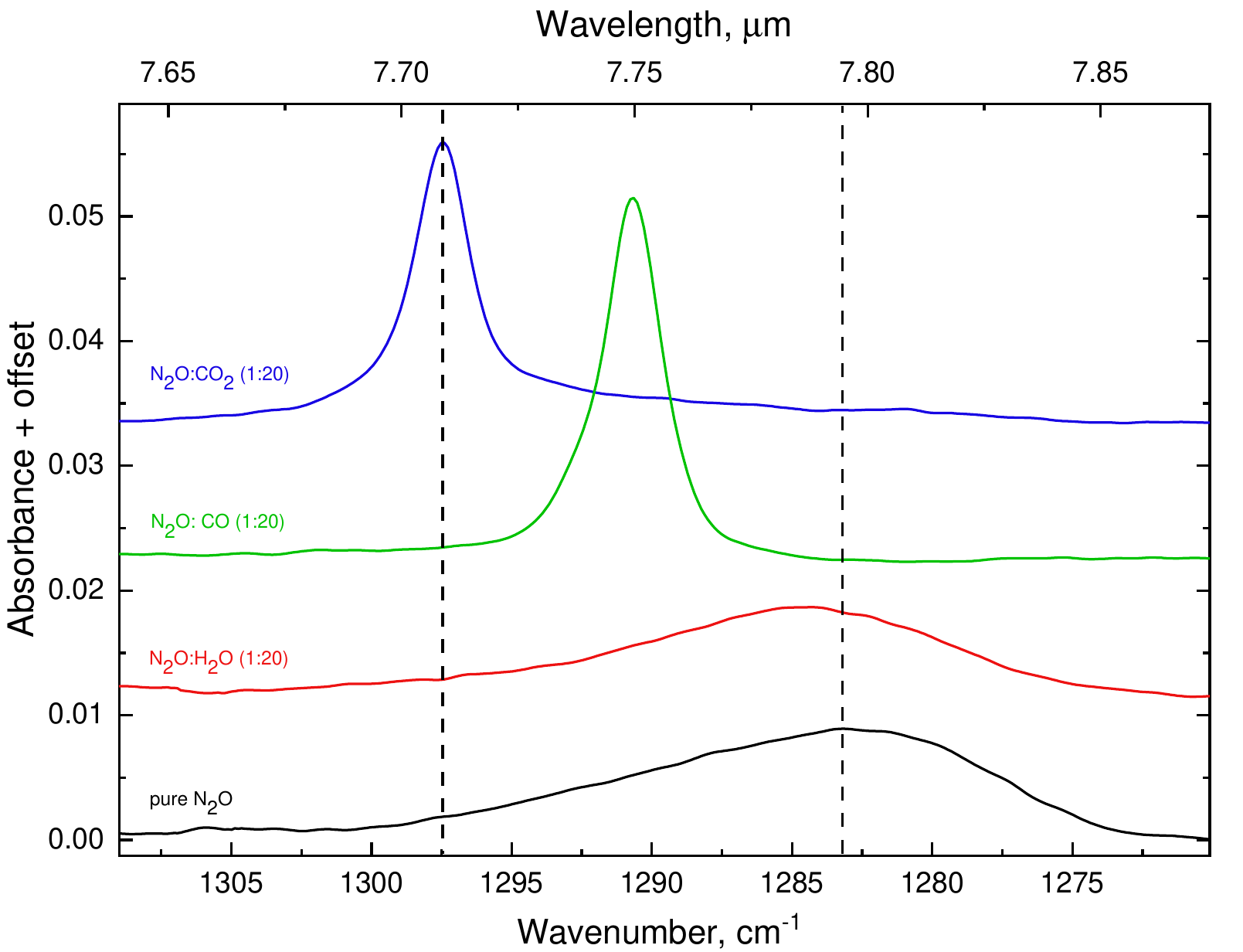}
    \caption{Laboratory spectra data of pure N$_2$O, N$_2$O~:~H$_2$O~=~1~:~20, N$_2$O~:~CO~=~1~:~20, N$_2$O~:~CO$_2$~=~1~:~20. All the mixtures have the same column density of both N$_2$O and a matrix molecule. The dashed lines mark the extreme peak positions of the N$_2$O features.}
    \label{N2O}
\end{figure}

Fig. \ref{N2O} shows N-O stretch region for pure N$_2$O and N$_2$O within the matrix of H$_2$O, CO or CO$_2$ at 10~K. The same amount of both N$_2$O and the matrix molecule was used in all experiments to maintain a 1~:~20 ratio in favor of the matrix molecule. The blue shift as much as 14.4~cm$^{-1}$ is observed in comparison to the peak position of the pure N$_2$O ice. The full set of IR ice spectra will be presented in the article Karteyeva et al. (in prep).

\begin{thebibliography}{}
\expandafter\ifx\csname natexlab\endcsname\relax\def\natexlab#1{#1}\fi
\providecommand{\url}[1]{\href{#1}{#1}}
\providecommand{\dodoi}[1]{doi:~\href{http://doi.org/#1}{\nolinkurl{#1}}}
\providecommand{\doeprint}[1]{\href{http://ascl.net/#1}{\nolinkurl{http://ascl.net/#1}}}
\providecommand{\doarXiv}[1]{\href{https://arxiv.org/abs/#1}{\nolinkurl{https://arxiv.org/abs/#1}}}

\bibitem[{Anderson \& Darling(1952)}]{AD_1952}
Anderson, T.~W., \& Darling, D.~A. 1952, The Annals of Mathematical Statistics, 23, 193 , \dodoi{10.1214/aoms/1177729437}

\bibitem[{Anscombe(1973)}]{Anscombe}
Anscombe, F.~J. 1973, The American Statistician, 27, 17, \dodoi{10.1080/00031305.1973.10478966}

\bibitem[{{Avni}(1976)}]{avni}
{Avni}, Y. 1976, \apj, 210, 642, \dodoi{10.1086/154870}

\bibitem[{Barman {et~al.}(2015)Barman, Konopacky, Macintosh, \& Marois}]{Barman_2015}
Barman, T.~S., Konopacky, Q.~M., Macintosh, B., \& Marois, C. 2015, ApJ, 804, 61, \dodoi{10.1088/0004-637X/804/1/61}

\bibitem[{{Bell} {et~al.}(2023){Bell}, {Welbanks}, {Schlawin}, {Line}, {Fortney}, {Greene}, {Ohno}, {Parmentier}, {Rauscher}, {Beatty}, {Mukherjee}, {Wiser}, {Boyer}, {Rieke}, \& {Stansberry}}]{2023Natur.623..709B}
{Bell}, T.~J., {Welbanks}, L., {Schlawin}, E., {et~al.} 2023, \nat, 623, 709, \dodoi{10.1038/s41586-023-06687-0}

\bibitem[{{Beuther} {et~al.}(2018){Beuther}, {Mottram}, {Ahmadi}, {Bosco}, {Linz}, {Henning}, {Klaassen}, {Winters}, {Maud}, {Kuiper}, {Semenov}, {Gieser}, {Peters}, {Urquhart}, {Pudritz}, {Ragan}, {Feng}, {Keto}, {Leurini}, {Cesaroni}, {Beltran}, {Palau}, {S{\'a}nchez-Monge}, {Galvan-Madrid}, {Zhang}, {Schilke}, {Wyrowski}, {Johnston}, {Longmore}, {Lumsden}, {Hoare}, {Menten}, \& {Csengeri}}]{2018A&A...617A.100B}
{Beuther}, H., {Mottram}, J.~C., {Ahmadi}, A., {et~al.} 2018, \aap, 617, A100, \dodoi{10.1051/0004-6361/201833021}

\bibitem[{{Beuther} {et~al.}(2023){Beuther}, {van Dishoeck}, {Tychoniec}, {Gieser}, {Kavanagh}, {Perotti}, {van Gelder}, {Klaassen}, {Caratti o Garatti}, {Francis}, {Rocha}, {Slavicinska}, {Ray}, {Justtanont}, {Linnartz}, {Waelkens}, {Colina}, {Greve}, {G{\"u}del}, {Henning}, {Lagage}, {Vandenbussche}, {{\"O}stlin}, \& {Wright}}]{Beuther2023}
{Beuther}, H., {van Dishoeck}, E.~F., {Tychoniec}, L., {et~al.} 2023, \aap, 673, A121, \dodoi{10.1051/0004-6361/202346167}

\bibitem[{Boogert {et~al.}(2015)Boogert, Gerakines, \& Whittet}]{boogert2015observations_Ch4}
Boogert, A. C.~A., Gerakines, P.~A., \& Whittet, D.~C. 2015, ARA\&A, 53, 541, \dodoi{10.1146/annurev-astro-082214-122348}

\bibitem[{{Boogert} {et~al.}(1998){Boogert}, {Helmich}, {van Dishoeck}, {et~al.}}]{1998A&A...336..352B}
{Boogert}, A.~C.~A., {Helmich}, F.~P., {van Dishoeck}, E.~F., {et~al.} 1998, A\&A, 336, 352.
\newblock \url{https://ui.adsabs.harvard.edu/abs/1998A&A...336..352B}

\bibitem[{{Boogert} {et~al.}(1997{\natexlab{a}}){Boogert}, {Schutte}, {Helmich}, {Tielens}, \& {Wooden}}]{Boogert1997}
{Boogert}, A.~C.~A., {Schutte}, W.~A., {Helmich}, F.~P., {Tielens}, A.~G.~G.~M., \& {Wooden}, D.~H. 1997{\natexlab{a}}, \aap, 317, 929.
\newblock \url{https://ui.adsabs.harvard.edu/abs/1997A&A...317..929B}

\bibitem[{{Boogert} {et~al.}(1997{\natexlab{b}}){Boogert}, {Schutte}, {Helmich}, {et~al.}}]{1997A&A...317..929B}
{Boogert}, A.~C.~A., {Schutte}, W.~A., {Helmich}, F.~P., {et~al.} 1997{\natexlab{b}}, A\&A, 317, 929.
\newblock \url{https://ui.adsabs.harvard.edu/abs/1997A&A...317..929B}

\bibitem[{Bouilloud {et~al.}(2015)Bouilloud, Fray, B{\'e}nilan, Cottin, Gazeau, \& Jolly}]{bouilloud2015bibliographic}
Bouilloud, M., Fray, N., B{\'e}nilan, Y., {et~al.} 2015, MNRAS, 451, 2145, \dodoi{10.1093/mnras/stv1021}

\bibitem[{{Cesaroni} {et~al.}(2019){Cesaroni}, {Beuther}, {Ahmadi}, {Beltr{\'a}n}, {Csengeri}, {Galv{\'a}n-Madrid}, {Gieser}, {Henning}, {Johnston}, {Klaassen}, {Kuiper}, {Leurini}, {Linz}, {Longmore}, {Lumsden}, {Maud}, {Moscadelli}, {Mottram}, {Palau}, {Peters}, {Pudritz}, {S{\'a}nchez-Monge}, {Schilke}, {Semenov}, {Suri}, {Urquhart}, {Winters}, {Zhang}, \& {Zinnecker}}]{Cesaroni2019}
{Cesaroni}, R., {Beuther}, H., {Ahmadi}, A., {et~al.} 2019, \aap, 627, A68, \dodoi{10.1051/0004-6361/201935506}

\bibitem[{Childs \& Bragg(1936)}]{1936}
Childs, W. H.~J., \& Bragg, W.~H. 1936, Proceedings of the Royal Society of London. Series A - Mathematical and Physical Sciences, 153, 555, \dodoi{10.1098/rspa.1936.0022}

\bibitem[{Cleveland(1979)}]{loess}
Cleveland, W.~S. 1979, Journal of the American Statistical Association, 74, 829, \dodoi{10.1080/01621459.1979.10481038}

\bibitem[{{Dartois} {et~al.}(2024){Dartois}, {Noble}, {Caselli}, {Fraser}, {Jim{\'e}nez-Serra}, {Mat{\'e}}, {McClure}, {Melnick}, {Pendleton}, {Shimonishi}, {Smith}, {Sturm}, {Taillard}, {Wakelam}, {Boogert}, {Drozdovskaya}, {Erkal}, {Harsono}, {Herrero}, {Ioppolo}, {Linnartz}, {McGuire}, {Perotti}, {Qasim}, \& {Rocha}}]{dartois2024spectroscopic}
{Dartois}, E., {Noble}, J.~A., {Caselli}, P., {et~al.} 2024, NatAs, 8, 359, \dodoi{10.1038/s41550-023-02155-x}

\bibitem[{{Dominik} {et~al.}(2021){Dominik}, {Min}, \& {Tazaki}}]{2021ascl.soft04010D}
{Dominik}, C., {Min}, M., \& {Tazaki}, R. 2021, Astrophysics Source Code Library, ascl:2104.010.
\newblock \doeprint{2104.010}

\bibitem[{{Drapatz} {et~al.}(1987){Drapatz}, {Larson}, \& {Davis}}]{drapatz1987comets}
{Drapatz}, S., {Larson}, H.~P., \& {Davis}, D.~S. 1987, \aap, 187, 497.
\newblock \url{https://ui.adsabs.harvard.edu/abs/1987A&A...187..497D}

\bibitem[{Emtiaz {et~al.}(2020)Emtiaz, Toriello, He, \& Vidali}]{doi:10.1021/acs.jpca.9b10643}
Emtiaz, S.~M., Toriello, F., He, J., \& Vidali, G. 2020, The Journal of Physical Chemistry A, 124, 552, \dodoi{10.1021/acs.jpca.9b10643}

\bibitem[{Fontani {et~al.}(2004)Fontani, Cesaroni, Testi, Walmsley, Molinari, Neri, Shepherd, Brand, Palla, \& Zhang}]{fontani2004iras}
Fontani, F., Cesaroni, R., Testi, L., {et~al.} 2004, A\&A, 414, 299, \dodoi{10.1051/0004-6361:20031623}

\bibitem[{{Francis} {et~al.}(2024){Francis}, {van Gelder}, {van Dishoeck}, {Gieser}, {Beuther}, {Tychoniec}, {Perotti}, {Caratti o Garatti}, {Kavanagh}, {Ray}, {Klaassen}, {Justtanont}, {Linnartz}, {Rocha}, {Slavicinska}, {G{\"u}del}, {Henning}, {Lagage}, \& {{\"O}stlin}}]{2024A&A_joys_gases}
{Francis}, L., {van Gelder}, M.~L., {van Dishoeck}, E.~F., {et~al.} 2024, \aap, 683, A249, \dodoi{10.1051/0004-6361/202348105}

\bibitem[{Gerakines \& Hudson(2015)}]{Gerakines_2015}
Gerakines, P.~A., \& Hudson, R.~L. 2015, ApJL, 805, L20, \dodoi{10.1088/2041-8205/805/2/L20}

\bibitem[{{Gieser} {et~al.}(2023){Gieser}, {Beuther}, {van Dishoeck}, {Francis}, {van Gelder}, {Tychoniec}, {Kavanagh}, {Perotti}, {Caratti o Garatti}, {Ray}, {Klaassen}, {Justtanont}, {Linnartz}, {Rocha}, {Slavicinska}, {Colina}, {G{\"u}del}, {Henning}, {Lagage}, {{\"O}stlin}, {Vandenbussche}, {Waelkens}, \& {Wright}}]{2023A&AGieser_gasclusters}
{Gieser}, C., {Beuther}, H., {van Dishoeck}, E.~F., {et~al.} 2023, \aap, 679, A108, \dodoi{10.1051/0004-6361/202347060}

\bibitem[{{Harada} \& {Herbst}(2008)}]{2008ApJ_CH4_chains}
{Harada}, N., \& {Herbst}, E. 2008, \apj, 685, 272, \dodoi{10.1086/590468}

\bibitem[{{Hassel} {et~al.}(2008){Hassel}, {Herbst}, \& {Garrod}}]{2008ApJ...681.1385H}
{Hassel}, G.~E., {Herbst}, E., \& {Garrod}, R.~T. 2008, \apj, 681, 1385, \dodoi{10.1086/588185}

\bibitem[{{Kobayashi} {et~al.}(2017){Kobayashi}, {Geppert}, {Carrasco}, {Holm}, {Mousis}, {Palumbo}, {Waite}, {Watanabe}, \& {Ziurys}}]{long_kobayashi}
{Kobayashi}, K., {Geppert}, W.~D., {Carrasco}, N., {et~al.} 2017, AsBio, 17, 786, \dodoi{10.1089/ast.2016.1492}

\bibitem[{{Krissansen-Totton} {et~al.}(2018){Krissansen-Totton}, {Garland}, {Irwin}, \& {Catling}}]{2018AJ....156..114K}
{Krissansen-Totton}, J., {Garland}, R., {Irwin}, P., \& {Catling}, D.~C. 2018, \aj, 156, 114, \dodoi{10.3847/1538-3881/aad564}

\bibitem[{{Labiano, A.} {et~al.}(2021){Labiano, A.}, {Argyriou, I.}, {Álvarez-Márquez, J.}, {Glasse, A.}, {Glauser, A.}, {Patapis, P.}, {Law, D.}, {Brandl, B. R.}, {Justtanont, K.}, {Lahuis, F.}, {Martínez-Galarza, J. R.}, {Mueller, M.}, {Noriega-Crespo, A.}, {Royer, P.}, {Shaughnessy, B.}, \& {Vandenbussche, B.}}]{Labiano}
{Labiano, A.}, {Argyriou, I.}, {Álvarez-Márquez, J.}, {et~al.} 2021, A\&A, 656, A57, \dodoi{10.1051/0004-6361/202140614}

\bibitem[{Lacy {et~al.}(1991)Lacy, Carr, Evans, Baas, Achtermann, \& Arens}]{lacy1991discovery}
Lacy, J., Carr, J., Evans, N.~J., {et~al.} 1991, Astrophysical Journal, Part 1 (ISSN 0004-637X), vol. 376, Aug. 1, 1991, p. 556-560. Research supported by Texas Advanced Research Program., 376, 556, \dodoi{10.1086/170304}

\bibitem[{Lei {et~al.}(2024)Lei, Feng, \& Fan}]{lei2024spatialdistributionrmch4}
Lei, L., Feng, L., \& Fan, Y.-Z. 2024, The Spatial Distribution of $\rm CH_4$ and $\rm CO_2$ Ice around Protostars IRAS 16253-2429 and IRAS 23385+6053.
\newblock \doarXiv{2409.04217}

\bibitem[{McClure {et~al.}(2023)McClure, Rocha, Pontoppidan, {et~al.}}]{a}
McClure, M., Rocha, W., Pontoppidan, K., {et~al.} 2023, NatAs, 7, 431, \dodoi{10.1038/s41550-022-01875-w}

\bibitem[{Mitchell \& Lora(2016)}]{mitchell2016climate}
Mitchell, J.~L., \& Lora, J.~M. 2016, AREPS, 44, 353, \dodoi{10.1146/annurev-earth-060115-012428}

\bibitem[{{Molinari} {et~al.}(2008){Molinari}, {Faustini}, {Testi}, {Pezzuto}, {Cesaroni}, \& {Brand}}]{Molinari2008}
{Molinari}, S., {Faustini}, F., {Testi}, L., {et~al.} 2008, \aap, 487, 1119, \dodoi{10.1051/0004-6361:200809821}

\bibitem[{{Molinari} {et~al.}(1998){Molinari}, {Testi}, {Brand}, {Cesaroni}, \& {Palla}}]{Molinari1998}
{Molinari}, S., {Testi}, L., {Brand}, J., {Cesaroni}, R., \& {Palla}, F. 1998, \apjl, 505, L39, \dodoi{10.1086/311591}

\bibitem[{{Molinari} {et~al.}(2002){Molinari}, {Testi}, {Rodr{\'\i}guez}, \& {Zhang}}]{Molinari2002}
{Molinari}, S., {Testi}, L., {Rodr{\'\i}guez}, L.~F., \& {Zhang}, Q. 2002, \apj, 570, 758, \dodoi{10.1086/339630}

\bibitem[{Mumma {et~al.}(1996)Mumma, DiSanti, Russo, Fomenkova, Magee-Sauer, Kaminski, \& Xie}]{doi:10.1126/science.272.5266.1310}
Mumma, M.~J., DiSanti, M.~A., Russo, N.~D., {et~al.} 1996, Science, 272, 1310, \dodoi{10.1126/science.272.5266.1310}

\bibitem[{{Nazari, P.} {et~al.}(2024){Nazari, P.}, {Rocha, W. R. M.}, {Rubinstein, A. E.}, {Slavicinska, K.}, {Rachid, M. G.}, {van Dishoeck, E. F.}, {Megeath, S. T.}, {Gutermuth, R.}, {Tyagi, H.}, {Brunken, N.}, {Narang, M.}, {Manoj, P.}, {Watson, D. M.}, {Evans, N. J.}, {Federman, S.}, {Muzerolle Page, J.}, {Anglada, G.}, {Beuther, H.}, {Klaassen, P.}, {Looney, L. W.}, {Osorio, M.}, {Stanke, T.}, \& {Yang, Y.-L.}}]{nazari2024hunting}
{Nazari, P.}, {Rocha, W. R. M.}, {Rubinstein, A. E.}, {et~al.} 2024, A\&A, 686, A71, \dodoi{10.1051/0004-6361/202348695}

\bibitem[{{Novozamsky} {et~al.}(2001){Novozamsky}, {Schutte}, \& {Keane}}]{2001A&A...379..588N}
{Novozamsky}, J.~H., {Schutte}, W.~A., \& {Keane}, J.~V. 2001, \aap, 379, 588, \dodoi{10.1051/0004-6361:20011332}

\bibitem[{Ozhiganov {et~al.}(2024)Ozhiganov, Medvedev, Karteyeva, Nakibov, Sapunova, Krushinsky, Stepanova, Tryastsina, Gorkovenko, Fedoseev, \& Vasyunin}]{Ozhiganov_2024}
Ozhiganov, M., Medvedev, M., Karteyeva, V., {et~al.} 2024, ApJL, 972, L10, \dodoi{10.3847/2041-8213/ad6d5c}

\bibitem[{Razali {et~al.}(2011)Razali, Wah, {et~al.}}]{ad_2nd}
Razali, N.~M., Wah, Y.~B., {et~al.} 2011, J. Stat. Model. Analytics, 2, 21

\bibitem[{{Reid Thompson} \& Sagan(1984)}]{REIDTHOMPSON1984236}
{Reid Thompson}, W., \& Sagan, C. 1984, Icarus, 60, 236, \dodoi{10.1016/0019-1035(84)90187-8}

\bibitem[{Rey {et~al.}(2013)Rey, Nikitin, \& Tyuterev}]{methane_lines}
Rey, M., Nikitin, A.~V., \& Tyuterev, V.~G. 2013, Phys. Chem. Chem. Phys., 15, 10049, \dodoi{10.1039/C3CP50275A}

\bibitem[{Rocha {et~al.}(2024)Rocha, McClure, Sturm, Beck, Smith, Dickinson, Sun, Egami, Boogert, Fraser, Dartois, Jimenez-Serra, Noble, Bergner, Caselli, Charnley, Chiar, Chu, Cooke, Crouzet, van Dishoeck, Drozdovskaya, Garrod, Harsono, Ioppolo, Jin, Jorgensen, Lamberts, Lis, Melnick, McGuire, Oberg, Palumbo, Pendleton, Perotti, Qasim, Shope, Urso, Viti, \& Linnartz}]{rocha2024iceinventoryprotostarced}
Rocha, W. R.~M., McClure, M.~K., Sturm, J.~A., {et~al.} 2024, Ice inventory towards the protostar Ced 110 IRS4 observed with the James Webb Space Telescope. Results from the ERS Ice Age program.
\newblock \doarXiv{2411.19651}

\bibitem[{{Rocha, W. R. M.} {et~al.}(2024){Rocha, W. R. M.}, {van Dishoeck, E. F.}, {Ressler, M. E.}, {et~al.}}]{JOYS}
{Rocha, W. R. M.}, {van Dishoeck, E. F.}, {Ressler, M. E.}, {et~al.} 2024, A\&A, 683, A124, \dodoi{10.1051/0004-6361/202348427}

\bibitem[{Sakai {et~al.}(2008)Sakai, Sakai, Hirota, \& Yamamoto}]{Sakai_2008}
Sakai, N., Sakai, T., Hirota, T., \& Yamamoto, S. 2008, ApJ, 672, 371, \dodoi{10.1086/523635}

\bibitem[{{Slavicinska} {et~al.}(2023){Slavicinska}, {Rachid, M. G.}, {Rocha, W. R. M.}, {Chuang, K.-J.}, {van Dishoeck, E. F.}, \& {Linnartz, H.}}]{Slavicinska_et_al._(2023)}
{Slavicinska}, K., {Rachid, M. G.}, {Rocha, W. R. M.}, {et~al.} 2023, A\&A, 677, A13, \dodoi{10.1051/0004-6361/202346996}

\bibitem[{Sromovsky {et~al.}(2011)Sromovsky, Fry, \& Kim}]{SROMOVSKY2011292}
Sromovsky, L., Fry, P., \& Kim, J. 2011, Icarus, 215, 292, \dodoi{https://doi.org/10.1016/j.icarus.2011.06.024}

\bibitem[{Stephens(1974)}]{stephens_1974}
Stephens, M.~A. 1974, Journal of the American Statistical Association, 69, 730.
\newblock \url{http://www.jstor.org/stable/2286009}

\bibitem[{{Swain} {et~al.}(2008){Swain}, {Vasisht}, \& {Tinetti}}]{2008Natur.452..329S}
{Swain}, M.~R., {Vasisht}, G., \& {Tinetti}, G. 2008, \nat, 452, 329, \dodoi{10.1038/nature06823}

\bibitem[{{Terwisscha van Scheltinga, J.} {et~al.}(2018){Terwisscha van Scheltinga, J.}, {Ligterink, N. F. W.}, {Boogert, A. C. A.}, {van Dishoeck, E. F.}, \& {Linnartz, H.}}]{spirt}
{Terwisscha van Scheltinga, J.}, {Ligterink, N. F. W.}, {Boogert, A. C. A.}, {van Dishoeck, E. F.}, \& {Linnartz, H.} 2018, A\&A, 611, A35, \dodoi{10.1051/0004-6361/201731998}

\bibitem[{Thompson {et~al.}(2022)Thompson, Krissansen-Totton, Wogan, Telus, \& Fortney}]{thompson2022case}
Thompson, M.~A., Krissansen-Totton, J., Wogan, N., Telus, M., \& Fortney, J.~J. 2022, Proceedings of the National Academy of Sciences, 119, e2117933119, \dodoi{10.1073/pnas.2117933119}

\bibitem[{Virtanen {et~al.}(2020)Virtanen, Gommers, Oliphant, Haberland, Reddy, Cournapeau, Burovski, Peterson, Weckesser, Bright, {van der Walt}, Brett, Wilson, Millman, Mayorov, Nelson, Jones, Kern, Larson, Carey, Polat, Feng, Moore, {VanderPlas}, Laxalde, Perktold, Cimrman, Henriksen, Quintero, Harris, Archibald, Ribeiro, Pedregosa, {van Mulbregt}, \& {SciPy 1.0 Contributors}}]{2020SciPy-NMeth}
Virtanen, P., Gommers, R., Oliphant, T.~E., {et~al.} 2020, Nature Methods, 17, 261, \dodoi{10.1038/s41592-019-0686-2}

\bibitem[{Öberg {et~al.}(2008)Öberg, Boogert, Pontoppidan, Blake, Evans, Lahuis, \& van Dishoeck}]{spitzer_oberg_2008}
Öberg, K.~I., Boogert, A. C.~A., Pontoppidan, K.~M., {et~al.} 2008, ApJ, 678, 1032, \dodoi{10.1086/533432}

\end{thebibliography}
\end{document}